\documentclass{aa}
\usepackage{txfonts}
\usepackage{graphicx}
\usepackage{float}
\usepackage{lscape}
\usepackage{natbib}
\usepackage{url}
\bibpunct{(}{)}{;}{a}{}{,}

\newcommand{\ang}{\mbox{${\rm \AA}$~}}
\newcommand{\Teff}{\mbox{$T_{\rm eff}$}}
\newcommand{\Lg}{\mbox{$\log\,g$}}

\newcommand{\Msol}{\mbox{$M_{\odot}$}}

\begin{document}

\title{Spectral analyses of eighteen hot H-deficient (pre-) white
  dwarfs from the Sloan Digital Sky Survey Data Release 4}
  \titlerunning{Spectral analyses of 18 hot H-deficient (pre-)
    white dwarfs} 
  \author{S.~D. H\"ugelmeyer\inst{1} \and S. Dreizler\inst{1} \and
  D. Homeier\inst{1} \and J.~Krzesi\'nski\inst{2,3} \and
  K. Werner\inst{4} \and A. Nitta\inst{5} \and S.~J.~Kleinman\inst{6}}

\institute{Institut f\"ur Astrophysik, Georg-August-Universit\"at
G\"ottingen, Friedrich-Hund-Platz 1, 37077 G\"ottingen, Germany \and
Apache Point Observatory, New Mexico State University, 2001 Apache
Point Road, P.O. Box 59, Sunspot, NM 88349, USA \and Mt. Suhora
Observatory, Cracow Pedagogical University, ul. Podchorazych 2, 30-084
Cracow, Poland \and Institut f\"ur Astronomie und Astrophysik,
Eberhard-Karls-Universit\"at T\"ubingen, Sand 1, 72076 T\"ubingen,
Germany \and Gemini Observatory, 670 N. A'Ohoku Place, Hilo, HI 96720, USA
\and Subaru Telescope, National Astronomical Observatory of Japan, 650
N. A'Ohoku Place, Hilo, HI 96720, USA}

\date{Received 17 January 2006 / Accepted 10 April 2006}
 
\abstract
{The Sloan Digital Sky Survey Data Release 4 has provided spectra of
  several new PG~1159 stars and DO white dwarfs. This increase in
  known hot H-deficient compact objects significantly improves the statistics
  and helps to investigate late stages of stellar evolution.} 
{From the optical SDSS spectra, effective temperatures and surface
  gravities are derived in order to place the observed objects in an
  evolutionary context. Especially the connection between PG~1159 stars
  and DO white dwarfs shall be investigated.} 
{Using our non-LTE model atmospheres and applying $\chi^2$-fitting
  techniques, we determine stellar parameters and their
  errors. We derive total stellar masses for the DO white dwarfs using
  model evolutionary tracks.} 
{We confirm three PG~1159 stars, with one showing ultra-high
  excitation ion features, and one sdO which we originally classified
  as a PG~1159 star. Additionally, we re-analysed the known PG~1159
  star, PG~1424+535, with our new models. Furthermore, we present
  the first spectral analyses of thirteen DO white dwarfs, {three of which}
  show M-star features in their spectra, while two display ultra-high
  excitation ion features.}
{} 

\maketitle

\section{Introduction}

White dwarfs (WDs) represent the final evolutionary stage for over
90\% of all stars (initial mass ${\rm M}_i < 8 \, {\rm
M}_{\small{\odot}}$). Due to very high mass loss at the tip of the
Asymptotic Giant Branch (AGB), WD progenitors lose part of their
envelope, forming a planetary nebula in their subsequent evolution. The
remaining core of the star rapidly evolves toward high effective
temperatures (\mbox{$\Teff > 100\,000$~K}). When \element{H}- and
\element{He}-shell burning cease, the star enters the WD cooling
sequence. The evolution of these post-AGB objects is separated into a
H-rich and H-deficient sequence with a ratio of about 4:1.

In their catalog of spectroscopically confirmed white dwarfs from the
Sloan Digital Sky Survey \citep[SDSS,][]{2000AJ....120.1579Y} Data
Release 4 \citep[DR4,][]{sdss...dr4}, \citet{2006wdc...dr4E} report
31 DO white dwarfs and 10 PG~1159 stars. Following our spectral
analyses of 16 hot \element{H}-deficient objects from the DR4 WD catalog
\citep[10 DOs, of which 9 were classified as DOs and one as DB+M, 5 PG~1159
stars, and one sdO;][]{2005A&A...442..309H} which were included in DR3
\citep{2005AJ....129.1755A}, we now extend this work to 13 DOs and 4
PG~1159 stars which are available only since DR4. The
{remaining} 9 DOs from the DR4 WD
catalog not analysed in this or our previous paper are either too
noisy or insufficiently flux calibrated for a reasonable spectral
analysis.

\subsection{PG~1159 stars}

PG~1159 stars are evolutionary transition objects between the hottest
post-AGB and WD phases. The prototype of this spectroscopic class,
PG~1159-035, was found in the Palomar Green (PG) survey
\citep{1986ApJS...61..305G}. It shows a spectrum without detectable
H-absorption lines. It is dominated by \ion{He}{ii} and highly ionised
carbon and oxygen lines. PG~1159 stars are characterised by a broad
absorption trough around 4670~\ang composed {of} \ion{He}{ii}
4686~\ang and several \ion{C}{iv} lines, suggesting high effective
temperatures. Spectral analyses {yield} $\Teff = 75\,000 -
200\,000$~K and gravities of $\Lg = 5.5 - 8.0$
\citep{1991A&A...244..437W,1994A&A...286..463D,1996aeu..conf..229W}.
{Of the 28 PG~1159 stars known prior to the SDSS, ten are
low-gravity \citep[subtype lgE, ][]{1992LNP...401..273W} stars,
placing them in the same Hertzsprung-Russell-Diagram (HRD) region as
the hot central stars of planetary nebulae (CSPNe), while the others
are more compact objects with surface gravities of WDs (subtype A or
E).} Due to their {rarity}, the majority of the known
PG~1159 stars were discovered in large surveys \citep[Palomar Green,
Hamburg Schmidt (HS), ][]{1995A&AS..111..195H}. The most
recent and only discovery within the last 10 years, besides those from
the SDSS \citep{2004A&A...424..657W}, was an object from the Hamburg
ESO (HE) survey \citep{1996A&AS..115..227W}. The SDSS thus offers a
new and relatively unique opportunity to increase the number of known
PG~1159 stars.

The star PG~1159-035 (= GW Vir) also defines a class of variable
stars. \citet{1979wdvd.coll..377M} discovered low-amplitude non-radial
$g$-mode pulsations in this object. About one third of the PG~1159 stars show this
variability which is thought to be driven by cyclic ionisation of
carbon and oxygen as suggested by \citet{1986HiA.....7..229C} and
\citet{1987fbs..conf..309S}. {Analyses of HST spectra of
spectroscopically identified PG~1159 stars indicate that there is a
likely relation between high carbon and oxygen abundances and pulsations}
\citep[][hereafter DH98]{1998A&A...334..618D}, which is corroborated
by theoretical calculations of \citet{2004ApJ...610..436Q}.

The region in the HRD occupied by the PG~1159 stars overlaps
with that of the DO white dwarfs. Therefore, it is assumed that
gravitational settling of the heavier elements in the atmosphere of
the PG~1159 stars leads them to transition towards DO white dwarfs.

\subsection{DO white dwarfs}

White dwarfs can be separated into two distinct spectroscopic classes,
DA and non-DA white dwarfs. The former show a pure hydrogen spectrum
and can be found over the entire WD cooling sequence. The latter fall
into three subclasses: DO (\mbox{$45\,000~{\rm K} < \Teff <
120\,000$~K}), DB (\mbox{$11\,000~{\rm K} < \Teff < 30\,000$~K}), and
DC / DQ / DZ white dwarfs (\mbox{$\Teff < 11\,000$~K}; pure continuum
/ carbon / metal lines present). The spectroscopic appearance of the
DO and DB subclasses is determined by the ionisation balance of
\ion{He}{i} and \ion{He}{ii}. DO white dwarfs show a pure \ion{He}{ii}
spectrum at the hot end and a mixed \ion{He}{i/ii} spectrum at the
cool end. The transition to the cooler DB dwarfs, characterised by
pure \ion{He}{i} spectra, is interrupted by the so-called \mbox{``DB
gap''} \citep{1986ApJ...309..241L}. In the HRD region of white dwarfs
with \mbox{$30\,000~{\rm K} < \Teff < 45\,000$~K}, no objects with
H-deficient atmospheres have been observed to date. However,
\citet{2006dbgapE} describe over 25 objects taken from DR4 data which
likely fall in that temperature region. These new objects might
increase our knowledge of the spectral evolution of \element{He}-rich
white dwarfs considerably.

Additional constraints on stellar parameters may result from the
analysis of DO white dwarfs in binary systems. Currently, three DOs
with late type companions (PG~0046+078, PG~0237+116, and HD~149499~B)
are known. The first two have M-star companions with only a minor
contribution to the combined spectrum
\citep{1996A&A...311L..17H,1985ApjS...58..379W}. The latter one is
accompanied by a much brighter K-star which complicates the analysis of the
optical spectrum of the DO \citep{1979MNRAS.187...17W}.

As in the case of PG~1159 stars, the most recent discovery of a DO
white dwarf comes from the HE survey \citep{2004A&A...424..657W}, while
the PG and HS surveys have contributed the majority of the 19 DOs known
prior to the SDSS.

\section{Spectral Analysis}
\label{sec:specan}

In order to analyse the DO and PG~1159 spectra \citep[SDSS
  observations for these objects are described
  in][]{2005A&A...442..309H}, we calculated homogeneous,
  plane-parallel, non-LTE model atmospheres with a code based on the
  Accelerated Lambda Iteration method \citep[][ and
  references therein]{2003ASPC..288...31W}. For these types of stars,
  it is necessary to account for non-LTE effects in our
  models, as shown by \citet{1996A&A...314..217D} for DO white dwarfs
  and by \citet{1991A&A...244..437W} for PG~1159 stars. To compare
  our synthetic spectra to the observed spectra, we normalise the
  observed spectra with third order polynomials fit through the
  continua. The continua are determined by the normalised theoretical spectra.
  Lineshifts due to radial velocities are taken into account
  by means of cross-correlation.  This comparison procedure is
  performed by an \texttt{IDL} code routine in order to guarantee
  consistent results. We used a $\chi^2$-statistic to derive best-fit
  models and computed 1-$\sigma$ errors following \citet{1986ApJ...305..740Z}.

\subsection{Spectral analysis of PG~1159 stars}

\begin{figure*}[ht!]
  \centering 
  \includegraphics[bb= 77 365 608 774,clip,width=18cm]{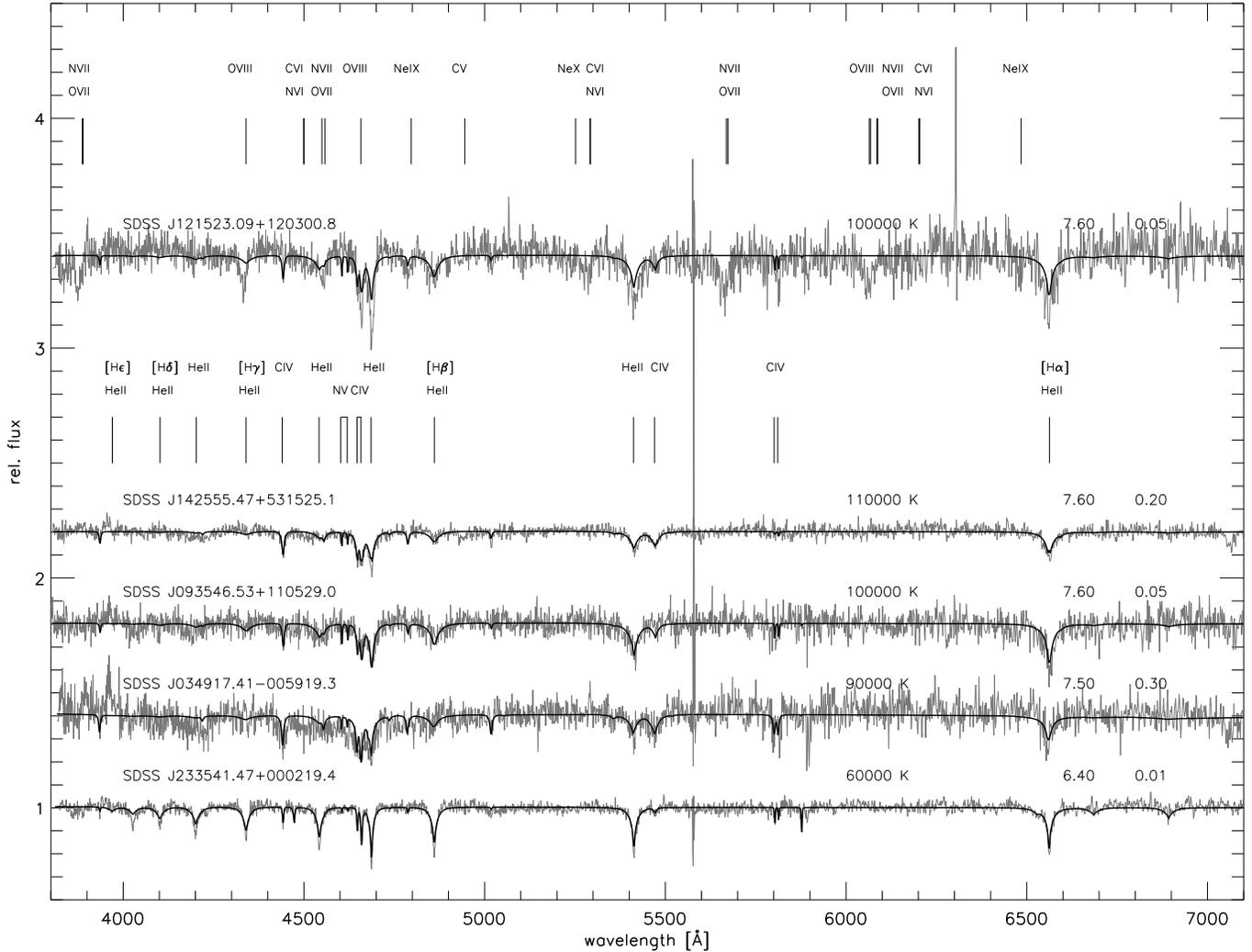}
  \caption{Normalised optical spectra (grey lines) of PG~1159 stars
    and the sdO star (bottom spectrum) along with model atmosphere fits (black
    lines). The top object shows ultra-high excitation ion
    features. Object names are printed on the left, with effective
    temperatures, logarithmic surface gravities, and carbon abundances
    (\element{C}/\element{He}), on the right.}
  \label{fig:pgcomp}
\end{figure*}

For the PG~1159 stars {in our sample}, we calculated model atmospheres
using detailed
\element{H}\,--\,\element{He}\,--\,\element{C}\,--\,\element{N}\,--\,\element{O}
model atoms. The model grid ranges from \mbox{$\Teff= 55\,000 -
150\,000$~K} and \mbox{$\Lg = 5.5 - 7.8$}. The abundances are fixed to
values $\element{He}/\element{H}=100$ and
$\element{C}/\element{He}=0.01$, $0.03$, $0.05$, $0.07$, $0.09$,
$0.10$, $0.11$, $0.20$, $0.30$, or $0.60$ by number. We have a nearly
complete model grid with $\Teff=55\,000 - 110\,000$~K in steps of
$5\,000$~K, $\Lg=6.4 - 7.8$ in steps of $0.2$~dex, and
$\element{C}/\element{He}=0.01-0.11$ in steps of $0.02$. Complete
coverage of the whole parameter space is not yet available due to the
high computational time required to compute the model
atmospheres. However, the majority of the analysed PG~1159 stars are
covered by our nearly complete model grid. Best-fit models for PG~1159
star candidates were calculated with an oxygen abundance following the
typical PG~1159 abundance-scaling ratio $\element{O}/\element{C}
\approx \element{C}/\element{He}$. However, variations in the oxygen
abundance do not {produce} a significant effect on the other stellar
parameters. The nitrogen abundance is kept fixed at
$\element{N}/\element{He}=0.01$ by number which is a typical upper
limit for PG~1159 stars. Analysed spectra of PG~1159 stars together
with their best-fit models are shown in Fig.~\ref{fig:pgcomp}.

\subsection{Spectral analysis of DO white dwarfs}
\label{sec:specan_do}

\begin{figure*}[ht!]
  \centering 
  \includegraphics[bb= 77 365 608 774,clip,width=18cm]{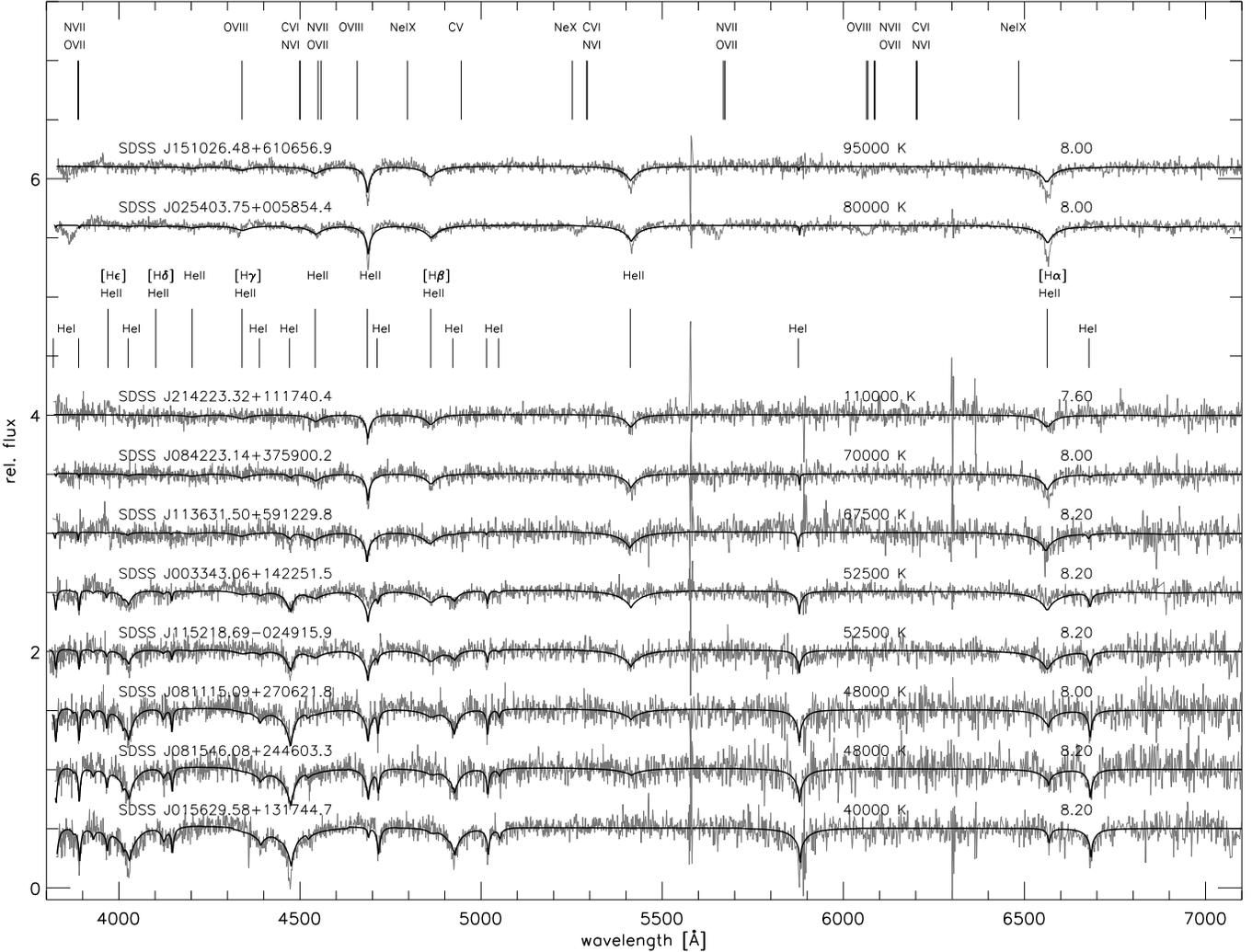}
  \caption{Normalised optical spectra (grey lines) of DO white dwarfs
  along with model atmosphere fits (black lines).  The two top objects show
  ultra-high excitation ion features. Object names are printed on the
  left, effective temperatures and logarithmic surface gravities on
  the right. The spectra of the hottest stars are dominated by
  \ion{He}{ii} absorption lines, while increasing line strengths of
  \ion{He}{i} are observed with decreasing temperature.}
  \label{fig:docomp}
\end{figure*}

Detailed \element{H}\,--\,\element{He} atomic models
\citep{1996A&A...314..217D} were used to calculate our model
atmospheres. The model grid ranges from
\mbox{$\Teff = 40\,000 - 120\,000$~K} in steps of $2\,500$~K,
\Lg\, ranges from 7.0 to 8.4 in intervals of 0.2 dex. The helium abundance
is fixed to $\element{He}/\element{H}=99$. The spectra of our analysed DOs,
together with their corresponding best-fit models, are shown in
Fig.~\ref{fig:docomp}.

\subsection{Analysis of spectra showing DO and M dwarf features}

In order to derive stellar parameters for both stars from the DO+M
star spectra, we used the DO model atmospheres previously described in
Section~\ref{sec:specan_do} along with theoretical LTE M-star atmospheres
calculated with the \texttt{PHOENIX} code \citep{ng-hot}. Their
construction follows the setup used for the preliminary GAIA grid
\citep{inesGAIA}, but has been updated to use \texttt{PHOENIX} version
14.1 and compute spectra at 0.2 {\AA} resolution. Major updates to the
input physics include the adaption of the revised solar composition of
\citet{2005ASPC..336...25A} for the chemical equilibrium abundances and a
correction to the partition sum for TiO, leading to $\sim$\,3-times
greater line strengths for this important absorber (Allard et al., in
prep.).

For the fit of both spectral components, a Planck spectrum was {first} fitted
to represent the Rayleigh-Jeans tail contribution of the DO in the red
part of the spectrum; then we selected a model spectrum based on an
initial fit of the spectral type of the cool companion and subtracted
it from the observation. Next, we used our $\chi^2$-fitting
routines to determine the best DO fit. This model was then subtracted
from the combined observed spectrum and the remaining cool star
spectrum was fitted to a grid of models spanning $\Teff = 2600-5000$~K
in steps of $100$~K and $\Lg= 3.5-5.5$ in steps of $0.5$ dex. The best
fit was determined by $\chi^2$-minimisation over the entire spectral
range; however we assigned a reduced weight to the regions from
6430\,--\,6560 \AA\ and 6720\,--\,6950\,{\AA}. In this part of the TiO
bands, the models reproduce the observations rather poorly, as has been
noted in previous analyses of M dwarfs
\citep[e.\,g.][]{MLTbinaries}. A reason for this mismatch is probably
superposed absorption due to CaH, for which no line opacity data are 
available in the \texttt{PHOENIX} version used for these models.
Despite these shortcomings we find acceptable fits for SDSS
J091621.83+052119.2 and SDSS J133633.22$-$013116.5 at
3600\,--\,3700\,K, with the fit quality rapidly degrading towards
higher and lower {\Teff}.  In light of the unknown systematic errors
due to the incomplete molecular data, flux calibration errors, and
possible interstellar reddening, we conservatively estimate an
uncertainty in {\Teff} of 100\,K. The much fainter spectrum of the
companion to SDSS J075606.36+421613.0 is consistent with temperatures
from at least 3000\,--\,3600\,K.  In all cases, the spectra show very
little sensitivity to gravity within a range of $4.0 < \Lg < 5.5$. A
direct determination of the stellar radius is therefore not possible
from these data, but the fits are entirely consistent with the
assumption of main sequence stars and the companion radius can
therefore be determined from a standard mass-radius-relation as well
as from evolutionary tracks.  We did not find any improvement of the
fits using metallicities either higher or lower than solar. 

Finally, we subtracted the best-fit M-star model atmosphere from the
observed spectrum and used our standard $\chi^2$-fitting routines
again to refine \Teff\ and \Lg\ for the DO white dwarf.  Only minor
differences in effective temperatures and surface gravities compared
to the first DO model fit were derived. The three spectra showing DO
and M-star features together with combined DO and M-star models are
presented in Fig.~\ref{fig:dom}.

\begin{table*}[ht!]
  \caption{Photospheric parameters of our PG~1159 star sample. The
  \element{C}/\element{He} abundance ratio is given by number. Objects
  marked with an asterisk ($^\ast$) have been analysed in
  \citet{2005A&A...442..309H}. The PG~1159 stars which have
  been observed for pulsations do not show variations above the specified
  detection limit $\sigma$.}
  \label{table:1}
  \centering
  \begin{tabular}{l r @{\,$\pm$\,} l r @{\,$\pm$\,} l r @{\,$\pm$\,} l
    l}
    \hline\hline\noalign{\smallskip} Name & \multicolumn{2}{c}{\Teff}
    & \multicolumn{2}{c}{\Lg} &
    \multicolumn{2}{c}{\element{C}/\element{He}} & remarks\\ 
    SDSS J & \multicolumn{2}{c}{[kK]} & \multicolumn{2}{c}{[cgs]} &
    \multicolumn{2}{c}{} & \\
    \hline 
    001651.42$-$011329.3$^\ast$ & 120.0&2.3 & 5.50&0.49  &
    0.30&0.006 & \\
    102327.41+535258.7$^\ast$   & 110.0&1.6 & 7.60&0.03  &
    0.20&0.007 & non-variable ($\sigma=12$~mmag)\\
    142555.47+531525.1      & 110.0&0.9 & 7.60&0.01 &
    0.20&0.004 & known as PG~1424+535 (DH98) \\
    075540.94+400918.0$^\ast$   & 100.0&2.6   & 7.60&0.05 &
    0.03&0.006 & \\
    093546.53+110529.0          & 100.0&1.3 & 7.60&0.03 &
    0.05&0.002 &  non-variable ($\sigma=15$~mmag)\\
    121523.09+120300.8      & 100.0&0.8 & 7.60&0.02 &
    0.05&0.002 & shows ultra-high excitation ion absorption \\
    134341.88+670154.5$^\ast$   & 100.0&0.7 & 7.60&0.02 &
    0.05&0.001 & non-variable ($\sigma=9$~mmag)\\
    144734.12+572053.1$^\ast$   & 100.0&1.6 & 7.60&0.02 &
    0.05&0.002 & non-variable ($\sigma=10$~mmag)\\ 
    034917.41$-$005919.3        &  90.0&0.9 & 7.50&0.01 &
    0.30&0.019 & \\ 
    \hline 
    233541.47+000219.4     &  60.0&0.9 & 6.40&0.01 &
    0.01&0.001 & (sdO) known as WD~2333$-$002 \\
    110215.46+024034.2$^\ast$   &  55.0&0.8 & 6.40&0.02 &
    0.01&0.003 & (sdO) \\
    \hline
  \end{tabular}
\end{table*}

\section{Results and discussion}

After presenting ten DO white dwarfs and five PG~1159 stars in
\citet{2005A&A...442..309H}, we now extend our spectral analyses to
the recently published SDSS DR4 DOs and PG~1159 stars. We present
results from this work and \citet{2005A&A...442..309H} together in
Table~\ref{table:1} and Table~\ref{table:2}. Some atmospheric
parameters for the PG~1159 stars analysed in
\citet{2005A&A...442..309H} have changed after applying our
$\chi^2$-fitting routines to these objects. The statistical
errors provided by the $\chi^2$-analysis and presented in
Table~\ref{table:1} and Table~\ref{table:2} are very small, while
suspected systematic errors are of the order of the grid steps
described in Section~\ref{sec:specan}.

\subsection{PG~1159 stars}

We found six spectra from the SDSS DR4 WD catalog
\citep{2006wdc...dr4E} which we identified as PG~1159 stars, although
one (SDSS J233541.47+000219.4) turns out to be best fit as a sdO
rather than a PG~1159 star, even though strong \ion{C}{iv} lines are
present. Our model grid is not adopted for the analysis of sdO stars
and the results are therefore preliminary only. We note that the
\ion{He}{ii} line cores are not well matched with the current best-fit
model. This object, known as WD~2333$-$002, has been previously
classified by \citet{2005ApjS..156...47L} as a sdB/sdOB.

One of our six selected PG~1159 stars, SDSS J114635.23+001233.5, was
previously known as PG~1144+005. Since no new model atmospheres for
this hot object are available, the photospheric parameters
$\Teff=150\,000$~K, $\Lg=6.50$, and $\element{C}/\element{He}=0.50$
derived by \citet{1991A&A...247..476W} still hold and the spectrum is
not analysed in this work.

SDSS J121523.09+120300.8 shows features of ultra-high excitation ions
(uhei) and is the first PG~1159 star of that type discovered so
far. Lines of \ion{O}{vii} and \ion{N}{vii} around 3888~\AA, at
5673~\AA, and around 6086~\AA, as well as \ion{O}{viii} at 4340~\AA,
4658~\AA, and 6064/6068~\AA\ are clearly visible in the spectrum (see
Fig.~\ref{fig:pgcomp}). Absorption features of \ion{C}{vi},
\ion{N}{vi}, \ion{Ne}{ix}, and \ion{Ne}{x} described by
\citet{1995A&A...303L..53D}, \citet{1995A&A...293L..75W}, and
\citet{2004A&A...417.1093K} for DO white dwarfs, are also indicated in
Fig.~\ref{fig:pgcomp} and are only ambiguously identifiable due to
low signal-to-noise and spectral resolution. The observed \ion{He}{ii}
absorption lines are too deep to be fitted with any model. Due to the
lack of appropriate modelling of the uhei phenomenon, stellar
parameters are derived for this star from a best-fit model which is
determined without considering the uhei lines. It should be kept in
mind that significantly higher temperatures (\Teff$\,>\,500\,000$~K)
are necessary to excite the uhei lines visible in this spectrum
\citep{1995A&A...303L..53D}.

\begin{figure*}[ht!]
  \centering \includegraphics[bb= 76 370 536
  560,clip,width=18cm]{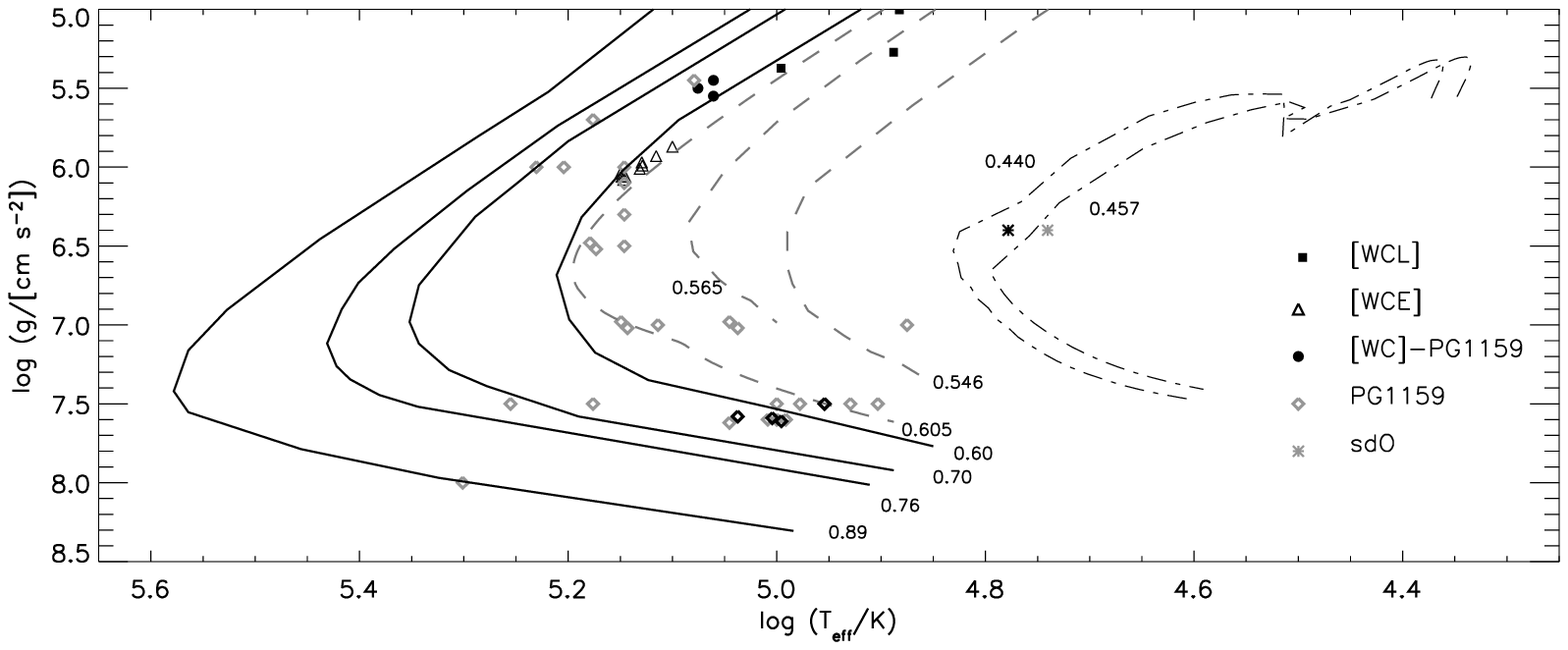} 
  \caption{Positions of the PG~1159 stars (black diamonds) analysed in
    this paper compared to evolutionary tracks from
    \citet{1995A&A...299..755B} and \citet{1983ApJ...272..708S} (dashed
    lines), and \citet*{1986ApJ...307..659W} (solid lines). The sdO star
    from this work (black asterisk) and the one from
    \citet{2005A&A...442..309H} are plotted together with tracks from
    \citet{1993ApJ...419..596D} (dash-dotted lines). Statistical error
    bars for all analysed objects are smaller than the symbols. The plot
    also includes previously known PG\,1159 stars and proposed progenitor
    objects (early and late Wolf-Rayet CSPNe:[WCE],
    [WCL], and [WC]-PG~1159). Labels: mass in M$_{\small{\odot}}$.}
  \label{fig:trackpg}
\end{figure*}

The remaining three PG~1159 star candidates are A subtype PG~1159
stars. SDSS J142555.47+531525.1 was already known as PG~1424+535 or
WD~1424+535 and has been analysed by DH98 and
\citet{1991A&A...244..437W} who derived
$\Teff=110\,000$~K/$100\,000$~K, respectively, $\Lg=7.00$, and
$\element{C}/\element{He}=0.30$ using UV and optical HST spectra. Our
fit provides the same effective temperature as DH98, but a higher
surface gravity ($\Lg=7.60$), and a lower carbon abundance
($\element{C}/\element{He}=0.20$) for the new SDSS spectrum of this
object. In our analysis we also fit the spectrum with the model used
by DH98, but we {obtained} a slightly lower $\chi^2$-value for our new
model. However, the models used are not very sensitive to carbon
abundance changes above $\element{C}/\element{He} \approx 0.10$ in the
optical.

The stellar parameters resulting from these spectral analyses are
listed in Table~\ref{table:1}. We have also included results from
time-resolved photometric observations of four candidates for pulsating
PG~1159 stars in Table~\ref{table:1}, obtained at the Calar Alto
Observatory 2.2~m telescope in March 2005. In Fig.~\ref{fig:trackpg},
positions of the four analysed PG~1159 stars and the one sdO star are
compared with evolutionary tracks as well as positions of previously
known PG~1159 stars including those from DR3
\citep{2005A&A...442..309H}.  The typical SDSS
PG~1159 star has approximately $\Teff=100\,000$~K, $\Lg=7.5$, and
$\element{C}/\element{He}=0.05-0.30$.

The position of SDSS J233541.47+000219.4 favours our classification of
it as a sdO star. It fits the evolutionary tracks of
\citet{1993ApJ...419..596D} within the tiny error bars very well and
therefore suggests an Extended Horizontal Branch rather than an AGB
history.

\subsection{DO white dwarfs}

Stellar parameters for 13 previously unanalysed DO white dwarfs are
presented in Table~\ref{table:2}. Three objects show M-star features
in their spectra and two display uhei lines \citep[classified
by][]{2004A&A...417.1093K}. Table~\ref{table:notana} lists the
DO white dwarfs from the DR4 WD catalog not analysed here -- mainly
because their spectra are too noisy.

The two uhei DO white dwarfs show absorption lines of \ion{O}{vii} and
\ion{N}{vii} around 3888~\AA, at 5673~\AA, and around 6086~\AA, as
well as \ion{O}{viii} at 4340~\AA, 4658~\AA, and
6064/6068~\AA. \ion{Ne}{ix} signatures are also present in SDSS
J025403.75+005854.4. The \ion{He}{ii} lines, especially the one at
6556~\AA, are very strong and cannot be fitted by our models. As
mentioned above, the effective temperature required to excite these ions to
a degree as detected in the two uhei DOs exceeds $500\,000$~K. A
massive post-AGB remnant of 1.2~\Msol\ can reach effective
temperatures up to $700\,000$~K on very short time scales
\citep{1970AcA....20...47P}, yet, \citet{1995A&A...293L..75W} argue
that uhei features in DO white dwarfs cannot be of photospheric origin
since \ion{He}{ii} lines would disappear completely at such high
temperatures. The hypothesis of optically thick and hot stellar winds
was proposed by the authors due to the triangular shape of the line
profile. Furthermore, the lines are blue shifted, which favours the
\begin{figure}[ht!]
  \centering \includegraphics[bb= 81 365 462
  684,clip,width=8.8cm]{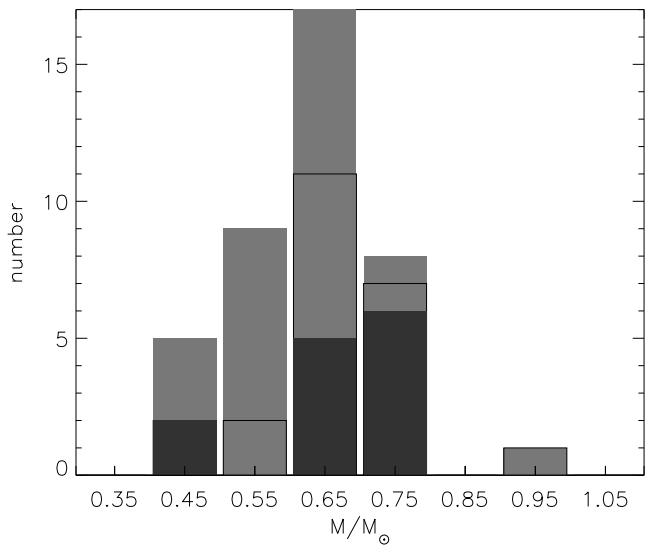}
  \caption{Mass distribution (bin\,=\,0.1\,\Msol) of all known DO
    white dwarfs. The black bars represent masses of stars from this
    work while the area under the black lines shows results from all
    the SDSS spectral analyses \citep[][ and this
    work]{2005A&A...442..309H}.}
  \label{fig:massdis}
\end{figure}
assumption of an expanding envelope. Neglecting this very hot
component in the spectral analyses, the uncertainties given by the
$\chi^2$-fitting procedure are likely to be lower than the systematic
errors due to incomplete modelling of the uhei phenomenon.
\begin{table}[ht!]
  \caption{Stellar parameters of our SDSS DO white dwarf sample. The
    three objects
    below the line show spectra with DO and M-star features. The letters
    A and B denote the DO and the M-star component, respectively. Objects
    marked with an asterisk ($^\ast$) have been previously analysed in
    \citet{2005A&A...442..309H}.}
  \label{table:2}
  \centering
  \begin{tabular}{l r @{\,$\pm$\,} l r @{\,$\pm$\,} l c}
    \hline\hline\noalign{\smallskip} Name & \multicolumn{2}{c}{\Teff}
    & \multicolumn{2}{c}{\Lg} & M \\ 
    SDSS J & \multicolumn{2}{c}{[kK]} & \multicolumn{2}{c}{[cgs]} &
    [\Msol] \\ 
    \hline 
    091433.61+581238.1$^\ast$   &120.0&1.7  & 8.00&0.02 & 0.75  \\
    204158.98+000325.4$^\ast$   &110.0&4.5    & 7.20&0.04 & 0.60  \\
    214223.32+111740.4          &110.0&2.4 & 7.60&0.05 & 0.60  \\
    154752.33+423210.9$^\ast$   & 100.0&1.6  & 7.60&0.06 & 0.59 \\
    151026.48+610656.9$^1$      & 95.0&0.8 & 8.00&0.04 & 0.71 \\
    084008.72+325114.6$^\ast$   & 85.0&1.1  & 8.40&0.08 & 0.90 \\
    025403.75+005854.4$^1$      & 80.0&0.5 & 8.00&0.06 & 0.68 \\
    155356.81+483228.6$^\ast$   & 75.0&1.2  & 8.00&0.09 & 0.68 \\
    084223.14+375900.2          & 70.0&0.3 & 8.00&0.06 & 0.68  \\
    140409.96+045739.9$^\ast$   & 70.0&0.5  & 8.00&0.04 & 0.68  \\
    113631.50+591229.8          & 67.5&1.6 & 8.20&0.13 & 0.78 \\
    131724.75+000237.4$^{\ast,2}$ &  62.5&0.5  & 7.80&0.03 & 0.58 \\
    034101.39+005353.0$^\ast$   & 55.0&0.4  & 8.00&0.05 & 0.65 \\
    003343.06+142251.5          & 52.5&0.2 & 8.20&0.01 & 0.76 \\
    115218.69$-$024915.9        & 52.5&0.3 & 8.20&0.02 & 0.76  \\
    034227.62$-$072213.2$^\ast$ & 50.0&0.2  & 8.00&0.03 & 0.65 \\
    081115.09+270621.8          & 48.0&0.3 & 8.00&0.04 & 0.64 \\
    113609.59+484318.9$^{\ast,3}$ &  48.0&0.1  & 8.00&0.04 & 0.64 \\
    081546.08+244603.3          & 48.0&0.2 & 8.20&0.01 & 0.76 \\
    015629.58+131744.7          & 40.0&0.2  & 8.20&0.01 & 0.75 \\
    \hline
    091621.83+052119.2 (A)   & 60.0&5.0 & 8.00&0.20 & 0.66 \\
    091621.83+052119.2 (B)   &  3.7&0.1 & 4.75&0.50 & 0.51 \\
    133633.22$-$013116.5\ A  & 52.5&0.5 & 7.60&0.06 & 0.48 \\
    133633.22$-$013116.5\ B  &  3.6&0.1 & 4.75&0.50 & 0.45 \\
    075606.36+421613.0\ A    & 52.5&0.8 & 7.40&0.05 & 0.42 \\
    075606.36+421613.0\ B    &  3.2&0.2 & 5.00&0.50 & 0.18 \\
    \hline
    \multicolumn{6}{l}{$^1$ shows ultra-high excitation ion absorption}\\
    \multicolumn{6}{l}{$^2$ previously known as HE~1314+0018
    \citep{2004A&A...424..657W}}\\
    \multicolumn{6}{l}{$^3$ previously known as WD~1133+489
    \citep{1997fbs..conf..303D}}\\
  \end{tabular}
\end{table}

SDSS J003343.06+142251.5 and SDSS J015629.58+131744.7 have been
classified by \citet{2004A&A...417.1093K} as DBAO and DBO,
respectively, while the former is listed as DO and the latter as DBO
in the DR4 WD catalog. Comparing the spectrum of the former object to
our DO models, we note that \ion{He}{ii} lines at 4686~\AA, 5412~\AA,
and 6563~\AA\ are stronger in the model than in the
observation. \ion{He}{i} is well fitted in the whole spectrum. In
contrast, \ion{He}{i} lines at 4026~\AA\ and 4471~\AA\ are stronger in
the observation than in the model for the latter. The
extension of the model grid to even lower model temperatures does not
improve the fit of the \ion{He}{i} lines since they are almost
saturated. The rest of the \ion{He}{i} lines in the model and the
observation match. SDSS J015629.58+131744.7 appears to be the coolest
DO known to date. The addition of small amounts of hydrogen did not lead
to a better fit for SDSS J003343.06+142251.5, so that we cannot explain
the weak \ion{He}{ii} lines by traces of hydrogen in the atmosphere.

\begin{figure*}[ht!]
  \centering 
  \includegraphics[bb= 74 372 534 560,clip,width=18cm]{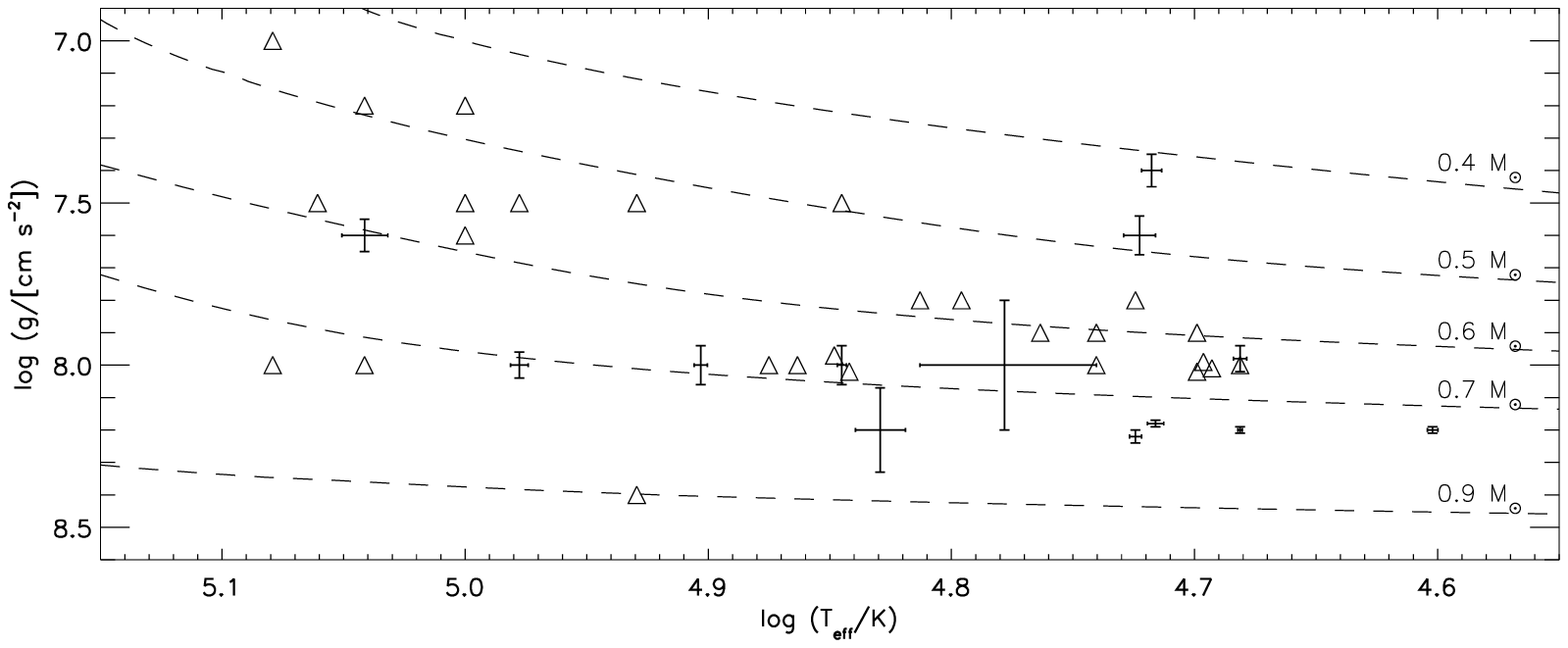}
  \caption{Positions of DO white dwarfs compared with evolutionary
  tracks from \citet{1995LNP...443...41W}. The triangles represent the
  27 hitherto known DOs
  \citep[see][]{1996A&A...314..217D,1997fbs..conf..303D,2004A&A...424..657W,2005A&A...442..309H}.}
  \label{fig:trackdo}
\end{figure*}

\begin{table}[ht!]
  \caption{DO white dwarfs from the DR4 WD catalog not analysed in
  this work due to noisy spectra or as denoted.}
  \label{table:notana}
  \centering
  \begin{tabular}{l l l}
    \hline \hline
    Name & Name \\
    SDSS J & SDSS J\\
    \hline
    011346.78+002828.7$^{1,2}$ & 140615.80+562725.9 \\
    074538.17+312205.3         & 150053.99+025143.2 \\
    081533.08+264646.4         & 205930.25$-$052848.9$^{3}$ \\ 
    083504.76+090111.7         & 213932.48+112611.2 \\
    130248.99$-$013309.5 & \\
    \hline 
    \multicolumn{2}{l}{$^1$ insufficiently flux calibrated}\\ 
    \multicolumn{2}{l}{$^2$ previously known as HS~0111+0012 \citep{1996A&A...314..217D}}\\
    \multicolumn{2}{l}{$^3$ dent in spectrum at $\sim 5800$ \AA}\\
  \end{tabular}
\end{table}

Masses for all objects are derived from our photospheric parameters
and evolutionary tracks from \citet{1995LNP...443...41W}. The thirteen
analysed DOs are plotted with these tracks in Fig.~\ref{fig:trackdo} and
the mass distribution for all DO stars is shown in
Fig.~\ref{fig:massdis}. As already mentioned in
\citet{2005A&A...442..309H}, DOs from the SDSS seem to have higher
masses than those analysed before the SDSS. We assume that this is
an effect introduced by the application of our automated fitting
routines which favour good line wing fits, thus steering the fits
towards higher surface gravities.

\subsection{Combined DO and M dwarf spectra}

\begin{figure*}[ht!]
   \centering \includegraphics[bb= 68 365 467
  718,clip,width=18cm]{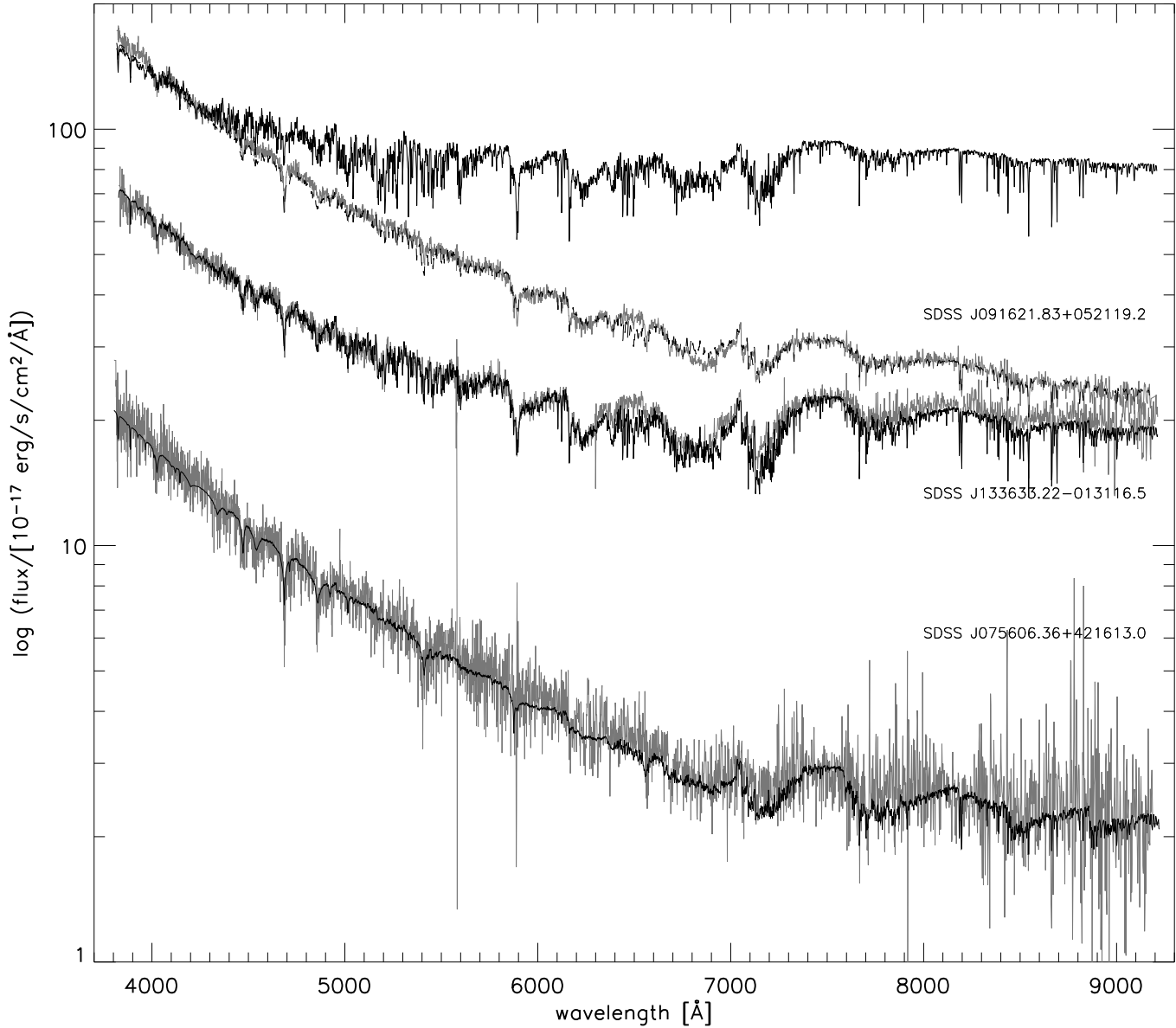}
  \caption{Observed spectra of two DO+M binaries (bottom two grey lines)
  and of one DO white dwarf and M-star pair which are coincidentally in the same
  line of sight. Combined model atmospheres for the two components are
  shown as black lines. The black dashed line plotted over SDSS
  J091621.83+052119.1 is the combination of the DO and M-star models
  where scaling is performed for different distances.}
  \label{fig:dom}
\end{figure*}

The three observed spectra that show both DO and M-star features, along
with combined DO and M-star model atmospheres are shown in
Fig.~\ref{fig:dom}. The flux of each model is multiplied by the
squared radius of the individual component. The radii are derived
from the surface gravity and mass which come from evolutionary tracks
from \citet{1995LNP...443...41W} for the DOs and
\citet{1998A&A...337..403b} for the M-stars. A constant factor which
accounts for our distance from the source is multiplied to DO and
M-star model fluxes. The total flux clearly fits SDSS
J133633.22$-$013116.5 and SDSS J075606.36+421613.0 within the
uncertainties, suggesting that the spectra come from binary
systems. SDSS J091621.83+052119.2, however, shows a deviation between
combined models and observation which cannot be explained by the
uncertainties {in} the stellar parameters. We therefore propose that
the spectrum originates from two {unrelated} stars on the same line of
sight. This is also corroborated by the {images of the objects, which
have ellipsoidal shapes and different photometric centres in different
colors}.\footnote{see
\url{http://cas.sdss.org/dr4/en/tools/explore/obj.asp?id=587732578300657729}}
A sum of DO and M-star models scaled \textit{individually} for distances fit
the observation well (dashed line in Fig.~\ref{fig:dom}) showing that
the model parameters were derived properly.

Stellar parameters for the M-star components are shown
in Table~\ref{table:2}. Our fitting routines found a best-fit model
with \Teff\,=55\,000~K for the DO part of SDSS J091621.83+052119.2
using the normalised spectrum. However, the flux distribution of the
observed spectrum favours an effective temperature of
\Teff\,=60\,000~K. We chose the latter value since M-star contributions
in the normalised spectrum (observation minus M-star model) were still
significant. Error estimates are 0.2 dex for \Lg\ and 5\,000~K in
\Teff\ because the continuum distribution is not properly fitted
with parameters which exceed the range given by these errors.

\section{Summary}

We have performed spectral analyses {of} 18 hot H-deficient (pre-)
white dwarfs that are new in Data Release 4 of the Sloan Digital Sky
Survey using our $\chi^2$-fitting routines. We confirm three PG~1159
stars, one of which shows lines of ultra-high excitation ions (uhei). A
re-analysis of PG~1424+535 with our $\chi^2$-programs using a SDSS
spectrum has provided a lower carbon abundance and a higher surface
gravity than found by \citet{1998A&A...334..618D}. Furthermore,
thirteen DO white dwarfs have been analysed, two of which show uhei
absorption and three of which have M-star features in their spectra. We
propose that two of the three latter objects are binary systems.  The
mass distribution of the DO white dwarfs from the SDSS is shifted towards
higher masses compared to the one constructed from analyses before the
SDSS.

\section{Acknowledgments}

We thank Daniel J. Eisenstein for helpful discussions and comments on
the DR4 WD catalog.

We would also like to thank F. Allard for instructive comments on molecular
opacities and helpful discussions. Model atmospheres used for this work
were calculated in part on the IBM pSeries of the Gesellschaft f\"ur
wissenschaftliche Datenverarbeitung mbH G\"ottingen (GWDG).

Funding for the SDSS and SDSS-II has been provided by the Alfred
P. Sloan Foundation, the Participating Institutions, the National
Science Foundation, the U.S. Department of Energy, the National
Aeronautics and Space Administration, the Japanese Monbukagakusho, the
Max Planck Society, and the Higher Education Funding Council for
England. The SDSS Web Site is http://www.sdss.org/.

The SDSS is managed by the Astrophysical Research Consortium for the
Participating Institutions. The Participating Institutions are the
American Museum of Natural History, Astrophysical Institute Potsdam,
University of Basel, Cambridge University, Case Western Reserve
University, University of Chicago, Drexel University, Fermilab, the
Institute for Advanced Study, the Japan Participation Group, Johns
Hopkins University, the Joint Institute for Nuclear Astrophysics, the
Kavli Institute for Particle Astrophysics and Cosmology, the Korean
Scientist Group, the Chinese Academy of Sciences (LAMOST), Los Alamos
National Laboratory, the Max-Planck-Institute for Astronomy (MPA), the
Max-Planck-Institute for Astrophysics (MPIA), New Mexico State
University, Ohio State University, University of Pittsburgh,
University of Portsmouth, Princeton University, the United States
Naval Observatory, and the University of Washington.

Photometric data for the search for pulsations in PG~1159 stars are
based on observations collected at the Centro Astron\'omico Hispano
Alem\'an (CAHA) at Calar Alto, operated jointly by the Max-Planck
Institut f\"ur Astronomie and the Instituto de Astrof\'isica de
Andaluc\'ia (CSIC).

\bibliographystyle{bibtex/aa}
\bibliography{dopg2}

\end{document}